\documentclass[conference]{IEEEtran}
\IEEEoverridecommandlockouts
\usepackage{cite}
\usepackage{amsmath,amssymb,amsfonts}
\usepackage{graphicx}
\usepackage{textcomp}
\usepackage{booktabs}
\usepackage{xcolor}
\usepackage{algorithm}
\usepackage{algpseudocode}
\usepackage{multirow}
\usepackage{hyperref}

\def\BibTeX{{\rm B\kern-.05em{\sc i\kern-.025em b}\kern-.08em
    T\kern-.1667em\lower.7ex\hbox{E}\kern-.125emX}}
\begin{document}

\title{Do Influencers Influence? -- Analyzing Players' Activity in an Online Multiplayer Game}
%\title{Follow Your Destiny -- Do Influencers Influence?}

%TODO: customize commands

%TODO: Add authors (Sigle Blinded)
\author{\IEEEauthorblockN{Enrica Loria}
\IEEEauthorblockA{
\textit{Fondazione Bruno Kessler}\\
%\textit{University of Trento}\\
Trento, Italy \\
eloria@fbk.eu}
\and
\IEEEauthorblockN{Johanna Pirker}
\IEEEauthorblockA{
\textit{Graz University of Technology}\\
Graz Austria\\
jpirker@iicm.edu}
\and
\IEEEauthorblockN{Anders Drachen}
\IEEEauthorblockA{
\textit{University of York}\\
York, UK \\
anders.drachen@york.ac.uk}
\and
\IEEEauthorblockN{Annapaola Marconi}
\IEEEauthorblockA{
\textit{Fondazione Bruno Kessler}\\
Trento, Italy \\
marconi@fbk.eu}
}

\maketitle

\begin{abstract}
In social and online media, influencers have traditionally been understood as highly visible individuals. Recent outcomes suggest that people are likely to mimic influencers' behavior, which can be exploited, for instance, in marketing strategies. Also in the Games User Research field, the interest in studying player social networks has emerged due to the heavy reliance on online influencers in marketing campaigns for games, as well as in keeping players engaged.
Despite the inherent value of those individuals, it is still difficult to identify influencers, as the definition of influencers is a debated topic.
Thus, how can we identify influencers, and are they indeed the individuals impacting others' behavior? In this work, we focus on influence in retention to verify whether central players impacted others' permanence in the game. We identified the central players in the social network built from the competitive player-vs-player (PvP) multiplayer (Crucible) matches in the online shooter Destiny. Then, we computed influence scores for each player evaluating the increase in similarity over time between two connected individuals. In this paper, we were able to show the first indications that the traditional metrics for influencers do not necessarily apply for games. On the contrary, we found that the group of central players was distinct from the group of influential players, defined as the individuals with the highest influence scores. Then, we provide an analysis of the two groups.
\end{abstract}

\begin{IEEEkeywords}
Social Network Analysis, Influencers, Player Behaviors, Game Analytics, Games User Research
\end{IEEEkeywords}

%8 pages including references and appendices
\section{Introduction}
The literature on online social networks~\cite{Kraut2011} suggests the existence of key individuals whose actions and behaviors are very impactful on other members of the community. Despite this influence or power that they exert on others can be of various forms~\cite{Goldhamer1939TypesStatus}, it can be understood as either a change or maintenance of a behavior conditioned by the influential individual~\cite{Goldhamer1939TypesStatus}. Thus, from a sociological perspective, we have an understanding of what influencers are: notable individuals that have an impact on others' behaviors. Due to this definition being very broad, formalizing it is a challenging task. As a consequence, several studies have developed their own approaches to model influence (e.g.,~\cite{Granovetter1985,Scripps2009,Goyal2010,Tan2010}). This problem is of interest in the Social Network Analysis (SNA) community since influencers have the power of disseminating messages and behaviors~\cite{Kraut2011}. For instance, they can be exploited in marketing strategies to promote certain products~\cite{Galeotti2009}. It follows that verifying whether influencers also have an impact on retention is extremely relevant in the game industry, and especially in multiplayer online games, which often rely on massive and loyal online communities to exist and survive. %Players' retention, as a result, is of vital importance. 
To date, influencers in the game community are mostly described as popular players that share their experiences on platforms~\cite{Churchill2016TheRelationships} or are active on social media~\cite{Wallner2019TweetingGame}.
However, one essential research question is still how we can find and identify influencers.
To asses whether influencers exist inside a game, and if they measurably affect peoples' behaviors, SNA can be used. A research study, in particular, has tackled the issue of identifying influencers through telemetry data~\cite{Canossa2019a}. In their work, influencers, who have been found to have an effect on players' retention, are intended as the central user in the network. Those nodes are important at a structural level, being at the center of many communications. Despite identifying influencers as central nodes in the network is valid and widely employed in the SNA literature, we are interested in analyzing whether those central players were actively influencing others, in terms of retention. Therefore, our main question is whether players influencing their neighbors' permanence in the game are those assuming a central position in the network. 
One might argue that a central player, having a wider net of connections, has a vast set of neighbors. As a consequence, the chances of being in contact with players long-retained in the game is higher. Besides, players more engaged in the game may be more attracted to players as engaged as they are (homophily~\cite{Lazarsfeld1964}). %We followed another line of thought, in which influencers are individuals that foster a change in their neighbors - i.e., people that they interact with. This type of influence is modeled differently~\cite{Shi2009}. 
As an alternative to observe the network from a structural perspective, the network can be analyzed semantically~\cite{Shi2009}. Towards this, temporal information is needed, since the analysis is performed on the evolution of the behaviors over time. Influence is evaluated as an increase in similarity among two nodes, since they first connect~\cite{Scripps2009,Crandall2008}.
In this work, we propose a methodology to compute influence, grounded in the SNA literature, where we measure influence as an increase in similarity~\cite{Scripps2009,Goyal2010}, mathematically computed as the cosine similarity of vector~\cite{Crandall2008}.

\subsection*{Research Questions}
The purpose of this work is to identify players' conditioning others' retention in the player-versus-player matches in the game \textit{Destiny}. We define these players as influencers, in that they \textit{influence} others' retention. To reduce wordiness, the term influence refers to the influence on other players' retention.  
%This is already methodology and not only research question 
%We employed two approaches to identify influencers: 1) a structural analysis on the network to detect central players, and: 2) a semantic analysis measuring influence as an increase in similarity over time. This resulted in two groups of players: \textbf{central} players, retrieved through the first approach, and \textbf{influential} players, retrieved through the second approach. For this purpose, we built a player social network from players' implicit interaction in the game. 
We investigated whether influencers defined as \textbf{central players} actually exert influence over other players, or whether \textbf{influential players}, defined as players' whose others' tend to mimic, are a distinct set of players. 
%We compared the two groups and investigated whether they affected their neighbors' behaviors in terms of retention - i.e., duration of the gameplay. Towards this, we defined a custom metric to measure the similarity of the gameplay's length of each node to its neighbors. This metric encloses information on the portion of neighbors who were retained in the game while the node remained active. We refer to this metric as \texttt{retention transfer}.

We propose the following research questions: 
\begin{itemize}
    \item[] \textit{\textbf{RQ1}.} Do all central players influence participation in other players?
    \item[] \textit{\textbf{RQ2}.} Are all influential players also central in the player social network?
\end{itemize}

\subsection*{Contribution}
In this work, we extended previous knowledge on influencers in games by employing a different approach to identify them. Instead of relying solely upon centrality measures~\cite{Canossa2019a}, we proposed an approach grounded from previous works in the social network analysis literature~\cite{Scripps2009}. 
Our algorithm measures how much players influence or are influenced by others in terms of retention. 
This study contributes to a better understanding of the whole network of players by defining a measure of influence, which not only can be used to detect influential users but also to measure how sensitive to neighborhood's influence certain nodes are. Besides, identifying influencers and players susceptible to influence is a knowledge that can be exploited by a matchmaking algorithm to intelligently inject relationships among users to prevent churn, which has concrete implications in the industry.

%Although the study focusing on data retrieved from the Crucible matches in Destiny, this study also aims at defining an approach - i.e., an algorithm - to measure users' influence. %Acquiring information on players' identity and their role in the community is particularly relevant in games for which the social component is one of the core mechanics, such as MOBAs and MMORPGs. 

\section{Related Work}
The work presented in this paper builds on previous work in two major domains: (1) social network analysis and (2) influencer analysis in social networks. 

\subsection*{Social Network Analysis} 
Applying Social Network Analysis (SNA) techniques to understand and investigate social structures, connections and interactions has become a commonplace strategy. Social networks are usually based on human interactions. The most common type of representation is a graph, where the nodes are the actors, and the edges consist of the interactions among the nodes~\cite{scott1988social}. %Despite the concept of social networks is not strictly linked to online communities, social media are often the main source of data. %It has become an commonly employed tool to understand social structures and dynamics in various application fields such as understanding social media networks such as Twitter or Facebook (e.g., ~\cite{ediger2010massive,kim2018social}), investigating information spread (e.g., ~\cite{hambrick2012six}), or spreading of disease.  
% due to page length I will remove references here... 
SNA has become a commonly employed tool to understand social structures and dynamics in various application fields such as understanding social media networks such as Twitter or Facebook, investigating information spread, or spreading of disease~\cite{wasserman1994social}. Also, for multi-user games, SNA is a vital tool to gain insights about the game and the players. 

%Social Network Analysis provides the tools for understanding the dynamics occurring in the network. Several interesting problems arise when studying the structural analysis of social networks, and several metrics have been defined. 
%Examples are: the distribution of the nodes' degrees; the existence of one or multiple connected components; and the ease and velocity of navigation of the network. 

\paragraph*{\textbf{SNA in Games}} %EXTEND
%THE IMPORTANCE OF STUDYING RELATIONSHIPS IN MP GAMES
Player interactions and relationships are a valuable source of information, which can be processed and modeled through social networks. Previous research on social connections and networks in games suggests that social connections and social interactions are essential motivational drivers for playing games~\cite{Williams2006}. %While,  The literature on online communities (e.g.,~\cite{Kraut2011}) suggests the presence of key elements contributing to the longevity of the community, the SNA in games is an underexplored research topic~\cite{Jia2015}. 
SNA in games has been used to identify relationships~\cite{Jia2015}, social roles of players~\cite{Ang2010}, analysis of groups and community~\cite{Ducheneaut2007,Ho2009}, the impact of social structures on performance and retention rate~\cite{Park2015,Pirker2018a}.     

%EXTERNAL COMMUNITIES AROUND GAMES
In the current literature, there is only little specific work on how player behaviors are related to player networks, social dynamics in games~\cite{Ducheneaut2007,Chen2008,Poor2015}), and the psychological aspect of the player and their in-game activities~\cite{SeifEl-Nasr2013}. As the study of the social dynamics in games is rare, the combination of network data and contextual data is almost nonexistent; exception made for the work of Rattinger et al.~\cite{Rattinger2016} and Schiller et al.~\cite{Schiller2018}. They studied how tools external to the game affect the group formation and the in-game interaction dynamics. Those studies contributed to the understanding of players by identifying and characterizing the groups formed and the roles that some players have within them (e.g., moderators, sherpas). Learning more about influential factors and influential players is essential. 

\subsection*{Influencers in Online Social Networks} 
Social networks are an essential tool to identify and understand influencers in online networks. In a social network, nodes often tend to resemble their neighbors. This happens either because similar individuals are driven towards one another, or because they mimic the behavior of some other individuals~\cite{Crandall2008}. The first phenomenon is called homophily~\cite{Lazarsfeld1964}, or selection, while the second is named social influence~\cite{Aggarwal2011}.

Influence is a widely studied topic in social network analysis, and yet there is no agreement on the definition of an influential person~\cite{Mahajan1990}. From state of the art, two types of influencers can be distinguished: (1) individuals affecting the spread of information or behavior~\cite{Shi2009}; and (2) individuals manifesting a particular combination of desirable properties, which span between expertise and position in the network~\cite{Freeman1978}. Many terms have been used to address those influential users. When they impact other behaviors, those individuals are referred to as opinion leaders~\cite{Easley2010}, innovators~\cite{Cacioppo2009}, key-players~\cite{Bonacich1987} and spreaders~\cite{Granovetter1973}. When they are well position and connected in the whole network, they are usually called celebrities~\cite{Newman2005}, evangelists~\cite{Backstrom2008} or experts~\cite{Granovetter1973}.

Using centrality measures to identify influencers has been proven to be a relevant approach~\cite{Granovetter1985,Bakshy2009}. More specifically, in- and out-degree, betweenness, eigenvector, and closeness are the more widely used metrics~\cite{Scripps2009}. Despite the fact that these measures are distinct, they are conceptually related~\cite{Rosenquist2010}. While, due to their definition, those metrics seem to be very aligned to the influencers of the second type, there is no trivial evidence that they are sufficient to identity influencers of the first type - i.e., influencing people behavior. Instead of being fundamental to keep the community connected, this specific kind of influencers foster similarity among the nodes, in that others tend to emulate them.

Many researchers have studied and modeled the concept of influence and its spreading throughout the network. Generally, in those works, influence is said to occur when B performs an action after A performed it. The probability of influence degree can be learned from a log of users' actions~\cite{Goyal2010}. A similar interpretation of influence, which is more tied to the individual's identity that their actions, is the study of the conditional probability that similarity increases from t-1 to t between two nodes that become linked at time t~\cite{Scripps2009}. Also, a combination of the two approaches is used, modeling both user attributes and actions over time. Tan et al. have used the latter methodology.~\cite{Tan2010} to compute the likelihood that the user also performs the action, which is increases when one's friends are performing said action. Various ways of measuring the influence of users have been analyzed~\cite{Bakshy2011,brown2011measuring,cha2010measuring}.
Studying the increase of similarity among users over time also allows modeling the idea of reinforcing influence when the interaction perpetuates. It is shown that similarity steadily increases even after the first interaction, although at a decreasing rate~\cite{Crandall2008}. Influence has also been used to differentiate between strong ties to weak ones~\cite{Xiang2010}. But most of this work has focused on online social networks, while work on influencers in games is still rare. 
%cosine similarity~\cite{Crandall2008}

\paragraph*{\textbf{Influencers in Games}}
While few recent works studied influencers in the online communities revolving around games (e.g., social media ~\cite{Wallner2019TweetingGame}, third-party websites ~\cite{Schiller2018} and other platforms~\cite{Churchill2016TheRelationships}) the study of influencers in gameplay is still neglected. A recent and notable exception is the work of Canossa et al.~\cite{Canossa2019a}, in which they analyzed the game Tom Clancy’s The Division (TCTD). They identified potential influencers as highly centralized players in the network. They also studied their properties to study how they differed from other users.

While previous work has focused on identifying and analyzing influencers and their properties with standard measures, it is still not clear if these standard measures (e.g., central players) have an impact on their neighbors and if the influential players are indeed these central players. In this work, we want to contribute to these research concepts by analyzing the impact of influential players and proposing methods on how to measure influence. 
%AT THE END of this section, important to summarize how we are building on and extending previous work, not only in games but in influencer studies/SNA in general. 

\section{Dataset}
%In the following section, we briefly introduce the game analyzed and the dataset built, as some statistical property of our data sample. 

\begin{table}[]
\caption{Temporal statistics of the Destiny dataset}
\centering
\begin{tabular}{@{}ll@{}}
\toprule
Property            & Value                                \\ \midrule
Observation Period  & 09 Sep 2014 -- 11 Aug 2015           \\
%Players             & 10 037                                \\
%Matches             & 26 120                             \\
\#Snapshots (days) & 336 \\
\#Snapshots (weeks) & 48   \\
\#Snapshots (months) &  $\sim$ 12 (4 weeks each) \\
 \bottomrule
\end{tabular}
\label{tab:statistics}
\end{table}

%\subsection*{Destiny}
\textit{Destiny} is a major commercial title, an online multiplayer first-person shooter (MMOFPS) video game developed by Bungie, released for multiple platforms. Despite being an FPS, the game also incorporates MMO elements and role-playing. Moreover, the game features a multiplayer "shared-world" environment with elements of role-playing games, e.g., character development. 
Players personify Guardians, which protect the Earth's last safe city from different alien races. Guardians are asked to revive a celestial being: the Traveler. During their journeys to different planets, they investigate and defeat the alien enemies to avoid humanity's destruction. 

Activities can either be player versus environment (PvE) and player versus player (PvP). The competition aspect is powerful in \textit{Destiny}. PvP matches follow objective-based modes, together with traditional deathmatch game modes. A considerable part of the gameplay consists in multiplayer fights in a restricted environment, which can be accessed through the Crucible. The Crucible is a hub for PvP content, taking place in instances separated from the main game world. In this study, we analyzed data from Crucible matches.

%TODO verify total number of players in the dataset
\subsection*{Data Overview and Feature Selection}
The dataset was generated from the collection of player-versus-player (PvP) Crucible matches in \textit{Destiny}, on a time window of 48 weeks  (from September 9, 2014, to August 11, 2015). To contain the computation time, from the whole population of about 3M of players, we selected a sample of 10k players and 26k matches. The sample was obtained by filtering out players that played for less than five weeks. As we will discuss later in detail, we built a dynamic network to measure influence. Before this, to choose how frequent the sequential snapshots of the network needed to be, we analyzed three-time granularity: days, weeks, and months. To ensure a minimum window bigger than one month (defined as four weeks) for every player, and thus to have a player present in at least two subsequent views of the network, we had to exclude players active in the game for less than five weeks. The statistical properties of the dataset are shown in Table \ref{tab:statistics}.

The main focus of this work is on the players' influence. Thus, we modeled players' retention through the following features: the number of matches they have participated in, the average time between the matches, the average seconds spent in each match, and the percentage of completed matches. 
For every node, we also computed a custom metric we named \texttt{retention transfer}, evaluating how much the node's neighbors were conditioned - either positively or negatively - by their retention. This evaluation metric will be used to verify whether the influencers identified have an impact on others' retention. The \texttt{retention transfer} lies in the interval $[0,\infty)$, where the lower the value, the more the node's neighbors emulated its retention. In other terms, values of retention transfer closer to 0 mean that the player's neighbors - on average - after the first encounter with the player, stayed in the game for as long the player stayed and left the game exactly when the player abandoned it. 
\begin{equation}
    rt_i = \frac{\sum_{j\in \mathcal{N}} |gameplay_i^t-gameplay_j^t|}{|\mathcal{N}|}
\end{equation}
where $\mathcal{N}$ is the set of $i$'s neighbors, $gameplay_i^t$ and  $gameplay_j^t$ are the length of the gameplay of the nodes i and j, respectively, after they first connected at time $t$.

%\begin{figure}[t]
%    \centering
%    {\includegraphics[width=0.6\columnwidth]{Figures/network_1y_components.png} }%
    %\qquad
%    {\includegraphics[width=0.6\columnwidth]{Figures/network_1y_communities.png} }%
    %\qquad
%    \caption{Overview of the (a) players social network components and (b) communities.}~\label{fig:network}
%\end{figure}

\section{Method: The Social Network of Destiny Crucible Players}
%In the following section, first, we describe the network structure and its properties. Then, we describe how we identified the set of central players and the set of influential players. We present the influence scores for the central players (RQ1) and how central influential players are (RQ2). 

\subsection*{Network Structure}
We built the player social network from the player-versus-players matches in \textit{Destiny}. %(Figure \ref{fig:network}). 
Each node in the network represents a player, and an edge exists if the two players were teammates in at least a match. The edges are weighted according to the number of matches the nodes shared. The network undirected since information on who initiated the match is unavailable. It should be noted that, since the opposing team cannot be chosen, the nodes (players) are connected if they participated in (at least) a match as teammates. 

We built both a static and dynamic network. The static network was used to compute the centrality measures, while the dynamic network was used to compute the influence scores. Table \ref{tab:net_stats} shows basic statistics of the static network.

\begin{table}[]
\caption{General Network Properties}
\centering
\begin{tabular}{@{}ll@{}}
\toprule
Property            & Value                                \\ \midrule
Nodes             & 10K                              \\
Edges             & 26K                              \\
Average Degree             & 4.60                             \\
Average Weighted Degree & 68.18                       \\
Diameter & 18                       \\
Modularity & 0.97                       \\
Connected Components & 1.5k                       \\
LCC & 42\% of the network                       \\
Second LCC & 41\% of the network                       \\
Average Clustering Coefficient & 0.55                       \\
Average Path Length & 6.61                      \\
\bottomrule
\end{tabular}
\label{tab:net_stats}
\end{table}

To analyze the evolution of players' behaviors, we used a dynamic player network. Despite the fact that the continuity of dynamic networks can be discretized by representing the network as a sequence of several snapshots, with no loss of information~\cite{Scripps2009}, the time between the snapshot is not universally defined. Thus, we investigated three levels of granularity (day, week, and month).
The choice of the granularity was led by the need for representing changes in players' participation behaviors over time. Thus, snapshots too far apart could have flattened the data, while increasing the frequency too much could have led to noisy data. To make a decision, we analyzed the players' participation levels over time. More specifically, we measured sudden changes in the behavior (as peaks) - e.g., oscillations between high and low level of activity. We also measured the tendency (slope) of their activity levelto measure the existence of gradual increases or decreases. Finally, we considered the degree of variability through the RSD (relative standard deviation).

\subsection*{Central Players}
First, we analyzed the network structurally, and thus we researched players that were well positioned and well connected. In particular, measuring how well a node is positioned in the network helps to identify the elements that hold the community together, and, hypothetically, have a more prominent role in it. The definition of influencers as central players is very controversial in terms of the centrality measures to be used. However, in the literature, the most robust measures are degree centrality, closeness centrality, betweenness centrality, and eigenvector centrality~\cite{Scripps2009}. To ease the comparison with previous works on influencers in games (i.e.,~\cite{Canossa2019a}), we also consider PageRank. Players were marked as central if they had high scores in the following centrality measures.
\begin{itemize}
    \item[1.] \textit{Degree Centrality:} number of connections the node has.
    \item[2.] \textit{Closeness Centrality:} nodes accessibility from others.
    \item[3.] \textit{Betweenness Centrality:} the number of the shortest path in which the node is involved.
    \item[4.] \textit{Eigenvector Centrality:} the number of important connections the node has. The importance of a connection is determined by the centrality value of the other node.
    \item[5.] \textit{PageRank:} the portion of players that can be accessed through direct links.
\end{itemize}

The distribution of all centrality measures is skewed to the left, with a long right tail. Closeness centrality is an exception, being bimodal with a high peak at 0.1 and a much lower peak at 1. Table \ref{tab:CM_distrib} show the distribution of the values.

For each one of the CM, we defined a threshold by considering the top 10\% of the distribution of the scores. We selected the central players as the individuals achieving scores higher than the thresholds for every centrality measure. This resulted in a sample of 51 players. For those central players, we found that the distribution of the \texttt{retention transfer} skewed away from 0, which is the desirable value (Figure \ref{fig:retention_distr}b). Besides, when considering the intersection of the players in the top 1\% and 0.1\% of each centrality measure, the sets were empty.

\begin{table}[]
\centering
\caption{Distribution of the centrality measures values}
\begin{tabular}{@{}lccccc@{}}
\toprule
\textbf{} & \multicolumn{5}{c}{\textbf{Distribution}}                                   \\ \midrule
          & \textit{min} & \textit{25\%} & \textit{50\%} & \textit{75\%} & \textit{max} \\
DC        & 1            & 2             & 4             & 7             & 92           \\
CC        & 0.08         & 0.14          & 0.16          & 0.17          & 1            \\
BC        & 0            & 0             & 523           & 8.9k          & 634k         \\
EC        & 0            & 0.005         & 0.01          & 0.03          & 1            \\
Pagerank  & 2.4e-5       & 6.6e-5        & 9.1e-5        & 1.1e-4        & 1.2e-3       \\ \bottomrule
\end{tabular}
\label{tab:CM_distrib}
\end{table}

\begin{algorithm}[t]
\caption{Computing the influence score for each edge.}\label{alg_inf}
\begin{algorithmic}[1]
\Procedure{InfluenceScore}{$E, X$}%\Comment{The g.c.d. of a and b}
\ForAll{$e(i,j) \in E$}
    \ForAll{$t \in \{t_1, ..., t_k\}$}
    \State $value = EdgeInfluence(X_i^{t-1}, X_i^{t}, X_j^{t-1}, X_j^{t})$
    \State $inf_i(e) = InfluenceAdj(value, w(e))$
    \State $inf_j(e) = -inf_i(e)$
    \EndFor
\EndFor
\EndProcedure
\end{algorithmic}
\end{algorithm}

\begin{table*}[htbp]
\centering
\caption{Players' participation stability throughut their gameplay}
\begin{tabular}{@{}llcccccccccccccc@{}}
\toprule
\multicolumn{1}{c}{\textbf{}} & \multicolumn{5}{c}{\textbf{Days}}                                           & \multicolumn{5}{c}{\textbf{Weeks}}                                          & \multicolumn{5}{c}{\textbf{Months}}                                         \\ \midrule
                              & \textit{min} & \textit{25\%} & \textit{50\%} & \textit{75\%} & \textit{max} & \textit{min} & \textit{25\%} & \textit{50\%} & \textit{75\%} & \textit{max} & \textit{min} & \textit{25\%} & \textit{50\%} & \textit{75\%} & \textit{max} \\
Number of peaks               & 0            & 0.75          & 1.5           & 3.5           & 75           & 0            & 0.25          & 0.75          & 1.75          & 15.5         & 0            & 0             & 0.5           & 0.75          & 4.25         \\
Slopes                        & -9k          & -34.1         & -2.2          & 19.8          & 17k          & -40k         & -303          & -34.9         & 177           & 36k          & -78k         & -1.5k         & -217          & 735           & 90k          \\
RSD (\%)                      & 0.6          & 52.1          & 63.9          & 74.3          & 180.8        & 0.51         & 46.5          & 60            & 72.1          & 158.1        & 0.51         & 39.7          & 53.7          & 67.4          & 135.24       \\ \bottomrule
\end{tabular}
\label{tab:granularity}
\end{table*}

\subsection*{Influential Players}
%To analyze the evolution of players' behaviors, we used a dynamic player network. Despite the fact that the continuity of dynamic networks can be discretized by representing the network as a sequence of several snapshots, with no loss of information~\cite{Scripps2009}, the time between the snapshot is not universally defined. Thus, we investigated three levels of granularity (day, week, and month). Weeks resulted in the more appropriate window of time. 
In this work, we defined influencers as players who affected the retention level of other players, after having interacted with them. Such definition is grounded on top of previous SNA works~\cite{Scripps2009,Crandall2008,Goyal2010,Tan2010}, describing influence within a connection - edge - as an increase in similarity from time t-1 to time t. Since we are interested in retention and participation, we computed the similarity on participation metrics - i.e., number of matches, the time between matches, completion rate, etc.

We first computed the influence score for each edge (Algorithm \ref{alg_inf}) through the \texttt{EdgeInfluence} function. The function returns the influence occurring on that edge for node i; the influence of node j is the additive inverse of $inf_i$. The similarity, and hence, the influence, is computed as follows. The influence score is the similarity of the two nodes when only one of the two changed behavior from the previous timeframe. The interpretation is that while one player maintains its behavior, the other mimics it. This value is then adjusted ($InfluenceAdj$ function) considering the weight of the edge - number of matches. The higher the number of matches, the stronger the influence is expected to be considered as such. The penalty scores follow a logarithmic distribution. 

Finally, we computed the influence for every node, according to the edge influence. It must be noted that for every edge we stored the influence.
We computed influence on the edges for all snapshots. Then, we computed the influence score of each node as the average influence they have on their neighbors\footnote{Further details and a Python implementation are available at \url{https://github.com/enrlor/sinfpy}, as the Pypi package.}

\begin{equation}
    Influence_i = \frac{\sum_{j : e(i,j) \in E}inf_i(e)}{\sum_{t}deg(i^t)}
\end{equation}

We defined influential players as individuals in the top 10\%, 1\% and 0.1\% of the influence distribution~\cite{Keller2003,WattsPeter2007}.

\begin{figure*}[t]
    \centering
    {\includegraphics[width=0.6\columnwidth]{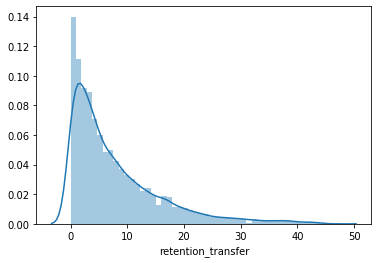} }%
    %\qquad
    {\includegraphics[width=0.6\columnwidth]{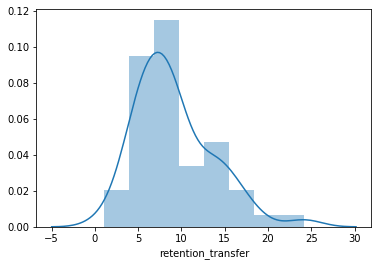} }%
    %\qquad
    {\includegraphics[width=0.6\columnwidth]{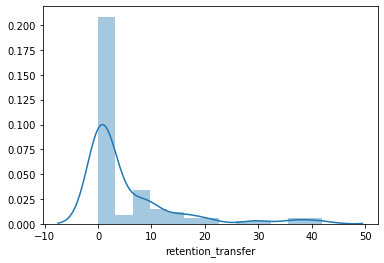} }%
    %\qquad
    \caption{Distribution of the retention transfer scores for all players (a), central players (b) and top 1\% influential players.}~\label{fig:retention_distr}
\end{figure*}

We found that the distribution of the \texttt{retention transfer} for the influential players peaked at the value of 0, with a long right tale (Figure \ref{fig:retention_distr}c).

\subsection*{Central vs Influential Players}
We compared the two groups of players - central and influential - retrieved with the two approaches described in the previous sections.

First, we observed that there was no intersection among the group of influential players and the group of central players. Then, we used the Mann-Whitney U test to verify whether there were any differences between central and influential players.
We found that the distribution of the influence score is considerably lower in central player, in comparison to the overall distribution of the influence score ($U = 1281, p = 2.3e-08, H_a$ = 'Influence score is higher for influential players than for central players'). Besides, the influence scores peaks at 0 for central players ($mean = -0.04, std - 0.11$), while it peaks at about 0.9 for influential players ($mean = 0.92, std - 0.03$).

We also compared the distribution of the five centrality measures - $H_a$ = 'CM is lower for influential players than for central players' in the Mann-Whitney U test, computed for each CM separately. 
For almost all of them, the influential players distinctly manifest lower values than central players (DC $U = 0, p = 3.3e-27$; BC  $U = 0, p = 2.5e-32$; CC  $U = 2450, p = 0.38$; EC  $U = 0, p = 3.5e-24$; Pagerank  $U = 0, p = 1.9e-24$). An exception is closeness centrality, for which we cannot say that influential players' values distribution is different than central players' (influential users having influence scores in the top 1\% and top 0.1\%).

Furthermore, we observed that, although having fewer connections, influential users' were involved in stronger links - i.e., the weights of the edges were higher. To verify that, we performed the Mann-Whitney U test on the average weighted degree of the nodes, with $H_a$ = 'Average WD is higher for influential players than for central players' ($U = 3k, p = 0.26$).

While influential users have very good scores in the \texttt{retention transfer value} (peak at 0), central players showed much higher values. Besides, the intersection of the two groups is empty.

We hypothesized the following scenario. Some central players could have influenced only part of their connections. Computing the node influence score as an aggregate of the edges' influence scores would have resulted in a loss of this information.
To verify that, we computed the standard deviation of the edge influence scores for every node. We found that central players have a significantly higher variability that influential players ($H_a$ = 'SD is lower for influential players than for central players', $U = 1281, p = 1.3e-08$).

It should be considered that (1) some players might not be influenceable, and (2) central players have a high degree.

\section{Results}
In the following section, we review the research questions we presented at the beginning of the paper.
We analyzed the gameplay data of the MMOFPS game \textit{Destiny}, retrieved from the Crucible matches. From the telemetry data, we built the implicit players' social network. We compared two ways to identify the influencers: a structural approach using centrality measures, and a semantic approach measuring nodes increase in similarity over time.

To decide the snapshot interval used to build the dynamic network we run preliminary analysis on players activity level. Results show that on average, the number of sudden changes in participation (peaks) is very low when we consider the months. As for weeks and days, the distribution is very similar. This means that this representation is more sensitive to changes, days even more so than weeks. When also considering the variability of those peaks, we saw similar behavior. The variability is lower for months, while it increases for weeks and days (the latter one again manifesting a similar distribution). 
We conducted a closer analysis of players lacking in extreme changes in participation. More specifically, we were interested in understanding whether those players assumed a constant behavior, or they increased (or decreased) their participation gradually. We measured the tendency (slope) of the level of participation. We saw that all distributions were very tall and skinny, with flat tails. However, the days' distribution is considerably taller than the other two, suggesting that players either changed their behavior drastically, or they maintained said behavior throughout the gameplay. We also studied the standard deviation of the slopes, which confirmed the finding. Finally, we measured the average relative standard deviation over the four participation features considered. We found once again that the variability is higher when structuring the network in days - the distribution is skewed towards the right. Also, in this case, the variability is lower when using months (skewed to the left), while weeks stay in between of the two. Thus, our final dynamic network was built of snapshots taken every week.

\subsubsection*{\textbf{RQ1. Do all central players influence participation in other players?}}
For each node, we computed five centrality measures: degree centrality, betweenness centrality, closeness centrality, eigenvector centrality, and PageRank. We marked as influencers the most central users, who had the highest scores in all the metrics. This led to a sample of 51 users, manifesting a poor \texttt{retention transfer} score. Thus, their neighbors showed a heterogeneous retention degree, both in case of long-term participation and in case of early abandonment. We also found that, generally, the influence scores of those users were very low, while the specific influence manifested on the edges resulted in a high variability. The elevated value of the standard deviations shows that those players were indeed influential, but such influence was effective only on a portion of their neighbors. This is explainable by the fact that, even in the real world, some individuals can be conditioned more easily than others despite the persuasive power of the person they are interacting with. Central users, having a broader range of connections, are in contact with players of different types, hence the variance in the results.

\subsubsection*{\textbf{RQ2. Are all influential players also central in the player social network?}}
For each node, we computed the influence score as the aggregate of the influence occurring on each edge they were involved in. The influence is a value in the range [-1;1], and is computed considering the evolution of participation behaviors of players over time. If a player tended to mimic another player's participation habits over time, with the latter being constant in their behavior, the first player was influenced by the second. We marked as influencers the players with the highest influence scores. This produced a set of 100 users, who, as the central users, presented values of the \texttt{retention score} very close to one (Figure \ref{fig:retention_distr}). We not only observed that the group of central players and the group of influential players were disjoint, but we also found that influential users had very low scores in all centrality measures. Thus, although they were in contact with fewer players - low degree centrality - they were the leading party of their group. In other terms, popularity (centrality) is not a synonym of influence. We also found that the average weighted degree is significantly higher for influential players than for central players. This can be interpreted as influential players generally having fewer but stronger relationships - i.e., more matches with the same players.

\section{Discussion}
Influencers are proven key users whose opinions have a substantial impact on the digital market~\cite{Goldhamer1939TypesStatus,Kraut2011}. 
Often they are identified as players strategically located in the network: central nodes~\cite{Granovetter1985,Bakshy2009}. In social medial, like Twitter, the contents are items that can be concretely shared, hence propagated physically in the community~\cite{Bakshy2011}. Thus, this privileged position makes those node disseminators~\cite{Bakshy2011}. Those highly visible users, reaching a broader audience, have an important role also in gamers communities~\cite{Wallner2019TweetingGame}. Previous works on groups formed around games, through third-party websites, show that the presence of specific figures (moderators) in the group correlated to the level of activity of its members~\cite{Schiller2018}. 

While some studies on online communities built around games exist (e.g.,~\cite{Wallner2019TweetingGame,Schiller2018}), the topic of influencers in games is still underexplored, a notable exception being Canossa's et al. work~\cite{Canossa2019a}. They analyzed telemetry data to investigate whether the influence also occurred in the actual gameplay. Their results showed that central players influenced others' retention. Players who connected with those individuals were more likely to stay in the game for longer. However, by definition, a central node has a favored position, resulting in an increased possibility to convey information to others~\cite{Hossain2007} in contrast to someone in the periphery of the network.\\
We further investigated how retention is encouraged (or hindered) by specific players and how they can be detected. Our algorithm is grounded in previous works in social network analysis~\cite{Scripps2009}, by adopting the definition of influence as an increase in similarity over time.
We extended Canossa et al.'s work~\cite{Canossa2019a} in two ways. First, we measured the actual influence exerted, as an increase in similarity on retention. Then, we allowed influence to be negative. In other words, in measuring whether a node tends to emulate another player's behavior, we include both an increase and a decrease of retention. \\
Studying players' interactions in the Crucible matches of \textit{Destiny}, similarly to Canossa's et al.~\cite{Canossa2019a} outcomes, we found that central users indeed affected others' retention. The influential users also seemed to have an impact on the retention of their neighbors.
Although one might argue that the custom metric used (\texttt{retention transfer}) is, to some extent, related to the definition of influence, comparing the length of the gameplay between players, this is not the case. While the retention transfer values consider the permanence of a user in the game, they ignore the participation features used, which are more vertical on the game itself. \\
Moreover, we found a dichotomy between central players and influential players. Not only we found that the centrality does not imply influence and vice versa, but we obtained two insights. First, influential players were involved in strong connections persistent over time, suggesting that strong influence can be reinforced. Second, we observed that despite central players not being selected as influential users, they exerted influence on a portion of their users. Therefore, influence is not an absolute property, but it also requires the other individual to be susceptible to be influenced.\\
In terms of the generalizability of the outcomes, we could hypothesize that something similar may occur when analyzing the whole gameplay. While it is true that the  Crucible matches vary from the general \textit{Destiny} gameplay, cooperation is still an active component in competitive matches, being a competition between teams. Thus, the social drivers that push players to play are similar. Nevertheless, influential players may be different individuals. The hypothesis is strengthened by the consistency of the results on how central players affect their neighbors in an online multiplayer game of a slightly different genre: TCTD, an RPG shooter~\cite{Canossa2019a}.\\
Our findings contribute to the knowledge of influencers in social communities and games by, first, presenting an algorithm to measure players' influence over time, and, second, by assessing whether central players' do exert influence on others' retention in Destiny \textit{Crucible} matches.

\paragraph*{\textbf{Industry Implications}} A better understanding of players' interaction in the game is fundamental, especially for games in which the multiplayer aspect is a core mechanics. Being aware of the dynamics occurring in the network may also highlight the company with what the players need and want. As previously observed by Canossa et al.~\cite{Canossa2019a} influencers have an impact on other players. They suggest exploiting those users to reach a more substantial part of the network. Thus, identifying influencers could have repercussions in the design of matchmaking algorithms. Players could also be matched depending on the type of influence the company is interested in. Not only might it help to detect players that have a positive effect on others, but also players that have a negative influence. Those players should be kept afar from users susceptible to influence, and maybe should be connected to strong positive influential individuals. The optimal use of social networks leads to higher sales and greater profits~\cite{Galeotti2009}.
Being aware of players' influence, in particular, whether it is positive or negative, can be used as additional information in matchmaking algorithms. This would help connecting players to reduce the possibility of churn by trying to exploit influencers' effect on retainment on players' more likely to mimic them - i.e., negative influence scores.

\paragraph*{\textbf{Limitations}} Despite the promising results, the study suffers from an obvious limitation: the specific nature of the network generated around Crucible matches, i.e. the PvP aspect of \emph{Destiny} , which could impact the generalizability of the findings. 
Due to the nature of the data, we have no information on other types of players' activities in \emph{Destiny}. This partial view of the whole gameplay experience constrains our findings to the context of inter-teams competition. This situation however mimicks most team-based competitive games, for example in esports, and we therefore expect similar outcomes in games that use analogous interaction mechanics - i.e., inter-team competition - since the social motivations that drive players to play are presumably the same.  In other words, we cannot say that influencers foster others' retention in the whole game of \emph{Destiny}, but we can say that influencers impact others' retention in competitive matches. 
In addition, while the literature suggests that different types of influence exist, in our work, we studied the influence on players' retention.

\section{Conclusions}
The importance of influencers derives from the impact that they have on people: their influence can potentially expand throughout the network. They can be exploited to attract (and retain) costumers, which is desirable in the marketing and, analogously, in the game industry. Nevertheless, few studies addressed this issue. The majority focused on describing the gamers community on social media platforms, as Twitter~\cite{Wallner2019TweetingGame} or formed through third-party websites~\cite{Schiller2018}, which might be different from influencers in the actual gameplay. In a pioneer work on influencers in games~\cite{Canossa2019a}, influencers have been defined as central players - i.e., users strategically positioned in the players' network. We investigated whether those central players were also influential users, defined as players exerting influence on their neighbors as an increase in similarity to their retention level in the game. Not only we found that central players are distinct from central players, but we also deepened our understanding of their role in the community. We observed (1) that influence is stronger when reinforced over time, and (2) that the status of influence is not absolute: players' can be influential only for a portion of their neighbors. These findings deepen our understanding of players involved in MMOFPS and can be used to inform matchmaking algorithms for games featuring inter-team competition.
In future works, we aim at researching the generalizability of our findings in other games, also belonging to different genres, as investigating the existence of types of influencers.

In summary, in the \textit{Destiny} Crucible matches, we found that \textit{influencers}, defined as central players, influence players retention in PvP matches to some extent, since some players are more susceptible than others to be influenced. Moreover, there are other influencers exerting influence, defined as in increase of similarity over time, that are involved in few strong connection, and thus, are not central or popular users.

\bibliographystyle{IEEEtran}
\bibliography{IEEEabrv,references}

\end{document}